# Structural and electronic properties of $V_2O_5$ and their tuning by doping with 3*d* elements – Modelling with DFT+*U* method and dispersion correction

A. Jovanović[1,2], A. S. Dobrota[1], L. D. Rafailović[2], S. V. Mentus[1,3], I. A. Pašti[1,4]*, B. Johansson,[4,5] N. V. Skorodumova[4,5]

[1]*University of Belgrade - Faculty of Physical Chemistry, Studentski trg 12-16, 11158 Belgrade, Serbia*

[2]*CEST Center of Electrochemical Surface Technology, Viktor-Kaplan Strasse 2, 2700 Wiener Neustadt, Austria*

[3]*Serbian Academy of Sciences and Arts, Knez Mihajlova 35, 11000 Belgrade, Serbia*

[4]*Department of Materials Science and Engineering, KTH - Royal Institute of Technology, Brinellvägen 23, 100 44 Stockholm, Sweden*

[5]*Department of Physics and Astronomy, Uppsala University, Box 516, 751 20 Uppsala, Sweden*

* **Corresponding author:** igor@ffh.bg.ac.rs

**Abstract**

New electrode materials for alkaline-ion batteries are a timely topic. Among many promising candidates, $V_2O_5$ is one of the most interesting cathode materials. While having very high theoretical capacity, in practice, its performance is hindered by low stability and poor conductivity. As regards theoretical descriptions of $V_2O_5$, common DFT-GGA calculations fail to reproduce both the electronic and crystal structure. While the band gap is underestimated, the interlayer spacing is overestimated as weak dispersion interactions are not properly described within GGA. Here we show that the combination of the DFT+$U$ method and semi-empirical D2 correction can compensate for the drawbacks of the GGA when it comes to the modelling of $V_2O_5$. When compared to common PBE calculations, with a modest increase of the computational cost, PBE+$U$+D2 fully reproduced the experimental band gap of $V_2O_5$, while the errors in the lattice parameters are only a few percent. Using the proposed PBE+$U$+D2 methodology we studied $V_2O_5$ doped with $3d$ elements (from Sc to Zn). We show that both the structural and electronic parameters are affected by doping. Most importantly, a significant increase of conductivity is expected upon doping, which is of great importance for the application of $V_2O_5$ in metal-ion batteries.

## 1. Introduction

Alkaline metal-ion batteries are one of the most commonly used and investigated electrochemical power sources of today. Among these, Li-ion batteries (LIBs) provide the highest energy density and the best cycle life. Although they show an appreciable performance, the search for new electrode materials for these systems continues.[1,2] Widely used electrodes, like $LiMn_2O_4$, $LiCoO_2$, and $LiNiO_2$, provide a noticeable performance and have found their application in commercial devices in spite of some drawbacks. Namely, the first commercial cathode material, $LiCoO_2$, although having a high theoretical capacity, can effectively deliver only around 150 mA h g$^{-1}$, in addition to its high price and pronounced toxicity.[3] In contrast to the mentioned materials, vanadium pentoxide ($V_2O_5$) attracted significant attention of the research community due to its high energy density, low cost, easy preparation, availability, and relatively safe use.[4–7] Li intercalation into $V_2O_5$ is followed by a series of first-order phase transitions, and reversible capacity corresponds to the phase $LiV_2O_5$. Further intercalation leads to irreversible transformations.[8] With the theoretical capacity of approx. 294 mA h g$^{-1}$ $V_2O_5$ outperforms commonly used cathode materials, making it a very promising cathode material for the next-generation of LIBs. In fact, the interest in $V_2O_5$ as an electrode material for rechargeable batteries (LIBs and other types of metal-ion batteries) has been revitalized due to the use of lithium metal as anode[3] and extensive search for cathode materials beyond LIBs.[3,9–12] The layered structure of $V_2O_5$ allows metal ion intercalation in-between $V_2O_5$ layers, causing the texture and morphology changes when metal ions are introduced into the structure.[8] While the electrode materials obtained from $V_2O_5$ show higher energy and power density, and are generally easier to prepare than conventional materials, the main drawback is the decrease in capacity during cycling, which is assumed to arise from the issues associated with low conductivity and material degradation.[13]

One of the strategies to overcome the problems related to stability and to improve the electrode performance of $V_2O_5$ is doping by various transition metals.[14,15] There are numerous reports in the literature showing the improved performance of doped $V_2O_5$ materials in rechargeable metal-ion batteries. As good examples, one can mention reports regarding the increased electronic conductivity of amorphous $V_2O_5$ upon doping with Ag, Cu, and Zn.[16–21] Also, an improved electrode performance was reported upon doping $V_2O_5$ with Mn [22]. In addition, versatile nanostructures of Cu-doped $V_2O_5$,[14,23] Fe-doped $V_2O_5$[24] and Cr-doped $V_2O_5$[25] were reported as cathodes for rechargeable metal-ion batteries, all witnessing an improved stability and better intercalation behavior of metal ions compared to pure $V_2O_5$.

Besides the large body of experimental work, $V_2O_5$ was also investigated theoretically, but to a lower extent. $V_2O_5$ is rather challenging for conventional Density Functional Theory (DFT), since due to correlation effects DFT methods underestimate the band gap and cannot account for dispersion interactions important for the description of $V_2O_5$.[26] In order to overcome the first problem, DFT+*U* approach [27,28] is often applied with satisfactory results.[29,30] The problem of the crystal structure description is addressed either by addition of (semi) empirical van der Waals (vdW) -terms to the Generalized Gradient Approximation (GGA) calculations[31] or using vdW-DF methods.[32] In particular Londero and Schröder [32,33] have shown that GGA-PBE (Perdew, Burke and Ernzerhof[34]) and GGA-PW91 (Perdew and Wang[35]) approaches, either using ultrasoft pseudopotentials or projector-augmented wave (PAW[36]) approach, significantly overestimate the interlayer spacing (around ~12%). The same authors have shown that the predicted structure is also very sensitive to the choice of the exchange-correlation functional within non-local van der Waals density functional methods: vdW-DF1[37,38] and vdW-DF2.[39] In fact, some of these functionals perform almost as poor as GGA-PBE in terms of structure description, while the energies of the interactions between $V_2O_5$ layers are estimated better. However, within the latter approach (the use of vdW-DF methods) the electronic structure of $V_2O_5$ is not described properly. To make the situation even more confusing, there are reports showing rather good agreement between the GGA results (PW91 results[29]) and experimental crystal structure. However, the agreement between theory and experiment became worse upon the addition of the on-site Coulomb correlation for the V $3d$ orbitals, in the attempt to describe the electronic structure better.[29]

The literature on the subject demonstrates that there is no consensus regarding the best way to treat $V_2O_5$ theoretically. One of the aims of the present work is to clarify this issue. As an adequate description of both electronic and crystal structures is required for a better understanding of materials performance, here we present a systematic analysis of $V_2O_5$ using plane wave DFT calculations. We show that a proper description of this material can be obtained by combining the DFT+*U* approach with the semi-empirical correction for the long range dispersion in the DFT+D2 formulation of Grimme.[40] Moreover, there is a significant number of experimental papers, which describe doped $V_2O_5$, demonstrating considerably improved performance, but a systematic theoretical analysis of the effects of doping on the properties of $V_2O_5$ is still lacking. Better understanding of the effects of doping can be of general interest to the battery research community and it can provide guidelines for designing novel electrode materials. Therefore, we also aim here at providing a general view on the effect of doping of $V_2O_5$ by $3d$ elements, focusing on the structure and electronic properties.

## 2. Computational details

The DFT calculations were performed using the GGA within Perdew–Burke–Ernzerhof parametrization for the exchange correlation functional,[34] applying the Quantum ESPRESSO (QE) *ab initio* package.[41] The pseudopotential method with ultrasoft pseudopotentials (USPP)[42], as implemented in QE, was used. The kinetic energy cutoff for the plane-wave basis set was 35 Ry, while the charge density cutoff was 16 times higher. Spin polarization was included in all the calculations. A simplified version of DFT+$U$ developed by Cococcioni and de Gironcoli[43] was used. The on-site Coulomb interaction was considered only for the vanadium $3d$ states and the value of $U$ was changed systematically between 2 eV and 6 eV. In order to account for the long range dispersion interactions the semi-empirical correction in the formulation of Grimme (DFT+D2) was applied.[40] The Brillouin zone was sampled using a $\Gamma$-centered k-point mesh, Gaussian smearing of 0.01 eV was applied to improve the convergence.

We doped $V_2O_5$ by $3d$ transition metals (denoted hereafter as M) both interstitially, between the layers of $V_2O_5$, and substitutionally, by replacing one of the V atoms in the simulation cell. The simulation cell was constructed as a (1×1×2) supercell of $V_2O_5$, containing two $V_2O_5$ layers along the $z$-direction of the cell. The doping was done in every second layer or in-between layers, resulting in the stoichiometry corresponding to $M_{0.25}V_{1.75}O_5$ (substitutional doping) or $M_{0.25}V_2O_5$ (interstitial doping). Preliminary calculations have shown that interstitial dopants prefer highly coordinated sites where $[MO_6]$ octahedral units are formed, which agrees with some previous experimental observations.[44] In the unit cell of pristine $\alpha$-$V_2O_5$ this roughly corresponds to the crystal coordinates $(0.5;0;z)$ where $z$ is around 0.5. In pristine $\alpha$-$V_2O_5$ (space group Pmmn) this corresponds to Wyckoff position 2b. Nevertheless, all the atoms were fully relaxed so the dopants could adjust their local environment. The concentration of dopants is somewhat larger than the ones usually considered experimentally, although in some cases even higher concentrations of dopants were used. Nevertheless, we are interested in the overall trends and consider them to be accounted properly using the same concentration of dopants for all systems. As will be discussed later on, for doped $V_2O_5$ we applied $U$ to the V 3d states only. Variable cell dynamics was used for structural optimization allowing both the cell and atomic positions to relax. Following the optimization, all the structures were recalculated in order to account for the change in basis set. Graphics presented in this work was made using either the VMD code [45] or VESTA.[46]

## 3. Results and discussion

*3.1. Pristine* $V_2O_5$

We first discuss the crystal structure of pristine $V_2O_5$ and the impact of the applied computational scheme on the obtained crystal structure parameters (Fig. 1). A clear trend in the calculated lattice parameters *versus* increasing *U* is seen: parameter *a* decreases, while parameters *b* and *c* increase. While the relative errors obtained for *a* and *b* are less than 3% with respect to the experimental values, that is typical for PBE, parameter *c*, corresponding to the interlayer spacing, is overestimated by more than 13% and the error increases with the value of *U* (Fig. 1). Similar relative errors for parameter *c* (11.2%) were previously reported by Kerber *et al.*,[31] who applied PBE combined with the PAW approach. The same authors were able to reduce the error down to 2.97% by adding the correction for dispersion interactions (with *a* and *b* fixed to the experimental values). Similarly, Londero and Schröder[33] reported parameter *c*, which was overestimated by 11.5% and 12% for PBE with PAW and USPP, respectively. The work of Ganduglia-Pirovano and Sauer,[47] using PW91 functional combined with PAW approach, overestimates *c* by 10.8%. All these results reflect a well-accepted fact that the interlayer interactions in bulk $V_2O_5$ are very weak and dispersive in nature.

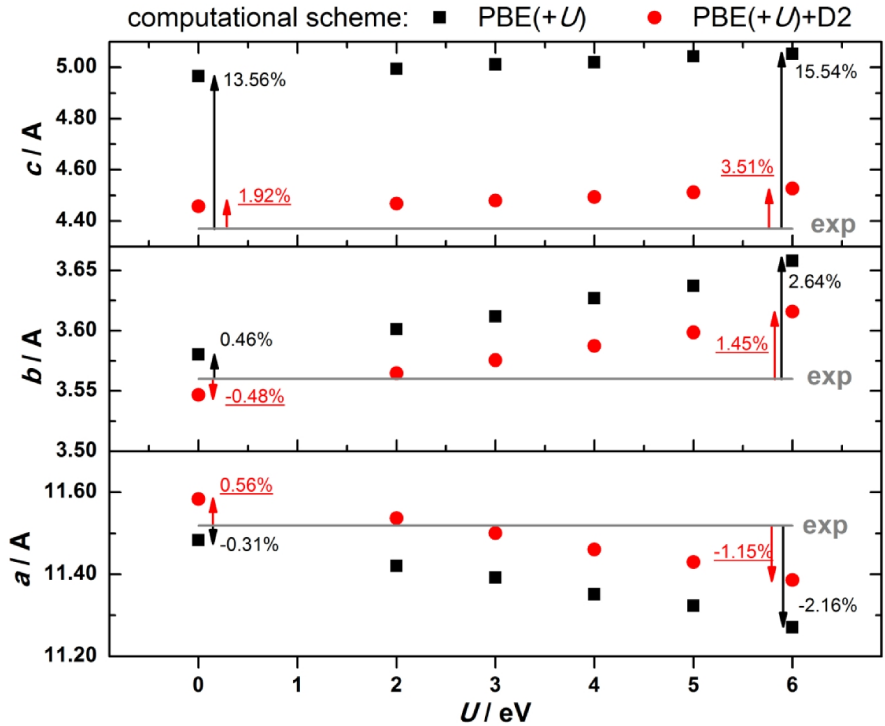

**Figure 1**. Dependence of the unit cell parameters of pristine $V_2O_5$ on the applied value of U, depending whether D2 correction was applied (circles) or no (squares). *U* = 0 is equivalent to plain PBE or PBE+D2. Indicated numbers give the relative errors (in %) of calculated lattice parameters with respect to the experimental values (underlined numbers are for PBE+*U*+D2 scheme).

Whereas the addition of on-site Coulombic $U$ applied to the vanadium $d$-states further increases the parameter $c$, the addition of the dispersion term compensates for this effect (Fig. 1). When PBE+D2 is applied, the error for $c$ is reduced down to 1.92%, i.e. 7 times compared to the PBE result. The impact of $U$ on the lattice parameters is noticeable and it remains the same irrespectively on the addition of D2 correction. Therefore, we conclude that the addition of the D2 correction to PBE+$U$ allows for the estimation of the lattice parameters with the relative error of a few percent only. Errors in the calculated lattice parameters are also translated into an overestimated unit cell volume (Fig. 2). It is clearly seen that the unit cell volume increases with $U$, with a relative error between 14% and 16% when dispersion interactions between the $V_2O_5$ layers are disregarded. This error arises almost exclusively from the error in parameter $c$ and it is efficiently removed upon the addition of D2 correction. Even for the highest value of $U$ the relative error found for the unit cell volume is below 4%, which we consider as an important improvement. When it comes to the battery applications of $V_2O_5$, the interlayer space provides the important diffusion paths for metal ions. Therefore, an incorrect interlayer distance would result in erroneous estimates of ion mobility in $V_2O_5$.

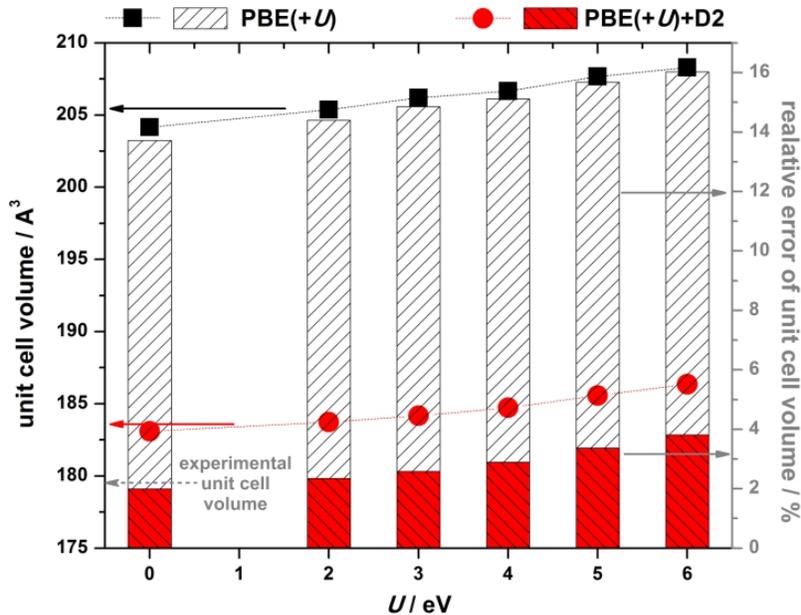

**Figure 2**. Unit cell volume of pristine bulk $V_2O_5$ depending on the value of $U$ for PBE+$U$ (squares) and PBE+$U$+D2 (circles). Relative errors are given by vertical bars. Dashed arrow gives the experimental value of the unit cell volume (179.53 $Å^3$). When $U$ = 0 eV, calculations are performed at PBE(+D2) level.

Let us now look at how well electronic structure is described within the PBE+$U$+D2 approach. Fig. 3 shows that the width of the band gap increases practically linearly with $U$, from ~1.6 eV (PBE and PBE+D2, $U$ = 0 eV) to ~2.3 eV (PBE+$U$ and PBE+$U$ +D2, $U$ = 6 eV). The width of the valence band (around 5.5 eV) depends weakly on $U$ and agrees with previously reported values.[29,48] The band gap is correctly estimated for higher values of $U$ ($U$ = 5 eV and $U$ = 6 eV). The experimentally determined values of the band gap of $V_2O_5$ are 2.0 eV[49] and 2.2 eV,[50] while Meyer *et al.*[51] have recently reported a bit larger value of 2.8 eV using the combination of ultraviolet, inverse and x-ray photoemission spectroscopy.

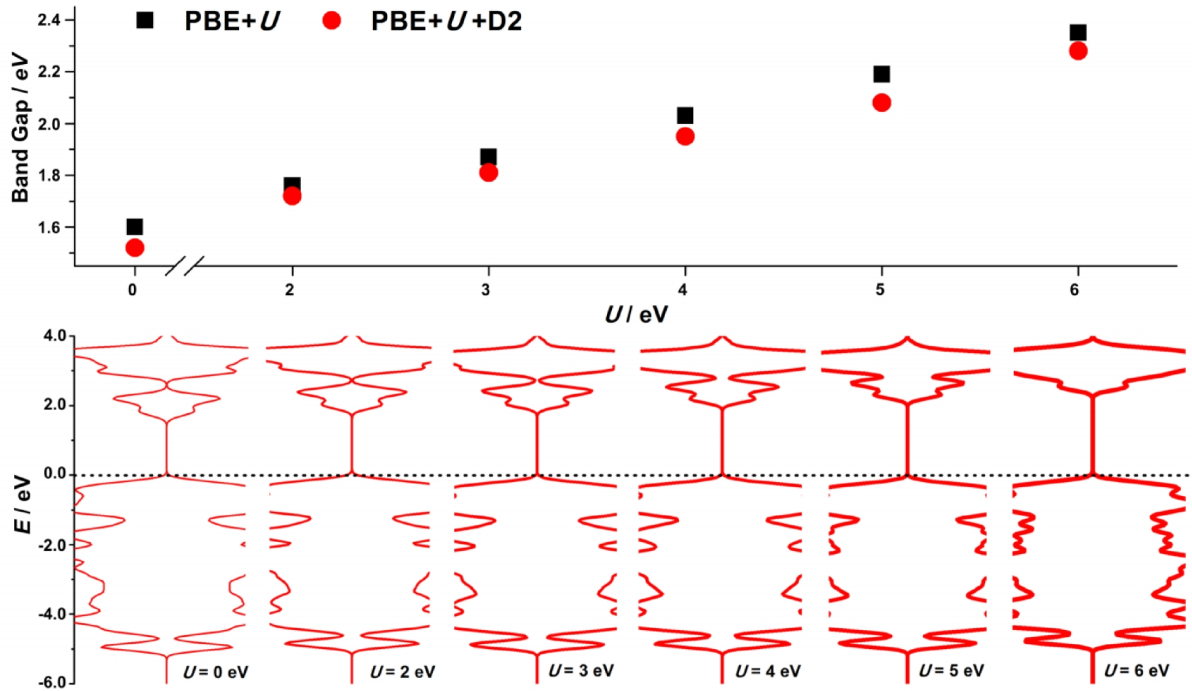

**Figure 3.** Calculated band gaps of pristine bulk $V_2O_5$ using PBE+$U$ (squares) and PBE+$U$+D2 (circles), top, and density of states (DOS) obtained using PBE+$U$+D2 approach. Top of the valence band is set to 0 eV.

The band gap determined with PBE is 1.6 eV, in agreement with previous work where the same level of theory has been used.[29,52] The gap calculated with PBE+$U$ shows better agreement with the experimental result, but it depends on the value of $U$ and the applied DFT+$U$ scheme. Most frequently, the approach of Dudarev *et al.*[28] is applied with $U_{eff}$ equal 3 eV or higher. Although the gap values close to the experimental ones have been found already for $U_{eff}$ = 3 eV,[29] in the literature there is no consensus about the optimum value of $U_{eff}$ (or the values of $U$ and $J$[27]). Nevertheless, considering the results presented in Fig. 3 and the range of the experimentally determined values of the band gap,[49–51] we chose to use $U$ = 6 eV hereafter, applied only to the 3$d$ states of V. It is also important to observe the impact of the D2 correction on the calculated band gap: PBE+$U$+D2 always gives slightly smaller band gaps than PBE+$U$. This is the consequence of the decrease of the interlayer spacing (Figs. 1 and 2).

While the effects of the D2 correction and the $U$ term on the lattice parameters are clear, it is also interesting to see how the chemical bonding within a single $V_2O_5$ layer is affected by them. The calculated bond lengths between vanadium and different oxygen atoms within the $V_2O_5$ layer (O(1), O(2) and O(3), Fig. 4) are summarized in Table 1.

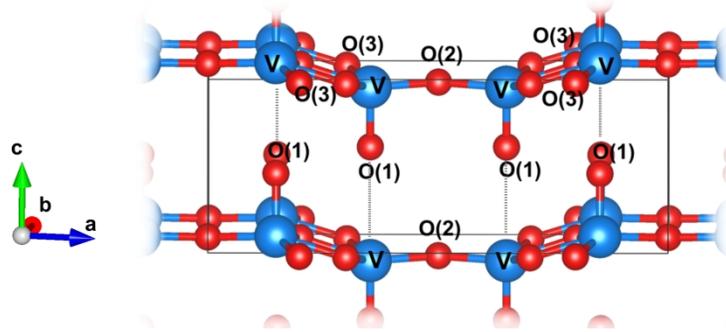

**Figure 4**. Notation of vanadium and oxygen atoms within a single $V_2O_5$ layer. Thin dashed lines show the interactions between the vanadyl O atoms of one $V_2O_5$ layer and vanadium atoms of a subsequent layer which are responsible for layer stacking in the bulk structure.

We note that relative differences between the calculated values and those determined in experiments are only a few percents. There is no single rule about the effect of addition of $U$ or D2 correction on the calculated bond lengths, but, to sum up, we do see that PBE+$U$+D2 in principle provides better agreement with experimental data than the PBE+$U$ scheme (Table 1). We ascribe this to the fact that strong chemical bonding within $V_2O_5$ layers is only weakly affected by the D2 correction, as discussed also previously for similar systems with strong chemical bonds.[53,54] It appears to be important to describe the interlayer interactions correctly. This is further supported by the fact that the contribution of dispersion interactions *per* $V_2O_5$ unit in our calculation is around 0.9 eV. This contribution depends only on the configuration of atoms in the simulation cell and it is affected by the value of $U$ in an indirect fashion through the modification of the lattice parameters (Fig. 1). This value closely matches the interlayer binding energies obtained using the vdW-DF methods.[33]

**Table 1**. Calculated vanadium-oxygen bond lengths (in Å) for different computational schemes, compared with the experimental data.

| | Bond lengths / Å | | |
|---|---|---|---|
| | V–O(1)* | V–O(2) | V–O(3) |
| PBE | 1.580 | 1.793 | 2.030/1.892 |
| PBE+D2 | 1.585 | 1.786 | 2.050/1.879 |
| PBE+$U$ | 1.576 | 1.804 | 1.982/1.924 |
| PBE+$U$+D2 | 1.579 | 1.794 | 1.999/1.908 |
| **Experiment*** | **1.57(5)** | **1.75(5)** | **2.03(8)/1.84(7)** |

*Ref. ([55])

Analyzing the computational cost of PBE+$U$+D2 calculations it can be concluded that the addition of $U$ increases the computational time by approx. 30-40%, whereas the addition of the D2 correction has no noticeable impact on the computational cost. In overall, we conclude that the drawbacks of the GGA in the case of $V_2O_5$ are successfully compensated for using the combination of DFT+$U$ and D2 methods. Hence, a relatively small increase of the computational costs for applying PBE+$U$+D2 is justified by the obtained results. We further apply the PBE+$U$+D2 scheme to study the structural and electronic changes in $V_2O_5$ upon doping with 3$d$ metals.

### 3.2. Doped $V_2O_5$

Having achieved the satisfactory description of the pristine system, we further study how the crystal and electronic structure of bulk $V_2O_5$ can be modified by doping with 3$d$ metals (from Sc to Zn). As mentioned earlier, we applied the PBE+$U$+D2 scheme for all the cases and applied on-site $U$ correction only to the 3$d$ states of vanadium. It should be noted that there is no unique strategy to select the value of $U$. Most often it is selected to fit to the experimental properties, which need to be reproduced to a satisfactory level.[56] In the recent overview by Capdevila-Cortada *et al.*[56] it was shown that a broad range of $U$ values was used for $d$-elements: from 1 eV to 10 eV.[56] As we are interested in the overall trends here, we chose not to apply $U$ to the $d$-states of dopants. For comparison, all the calculations were also repeated using PBE, PBE+$U$ and PBE+D2. We do not present the detailed results of those calculations here but their comparison allows us to conclude that the overall trends are reproduced by all of the approaches. The dopants were introduced either between the $V_2O_5$ layers or as a substitutional impurity in the layer (Fig. 5). We have also performed some additional calculations for the case of Mn- and Co-doped $V_2O_5$ where $U$ correction was systematically applied to Mn and Co $3d$-states as well. The results are presented in Supplementary Information and show that the main conclusions are not affected by sensible values of $U$.

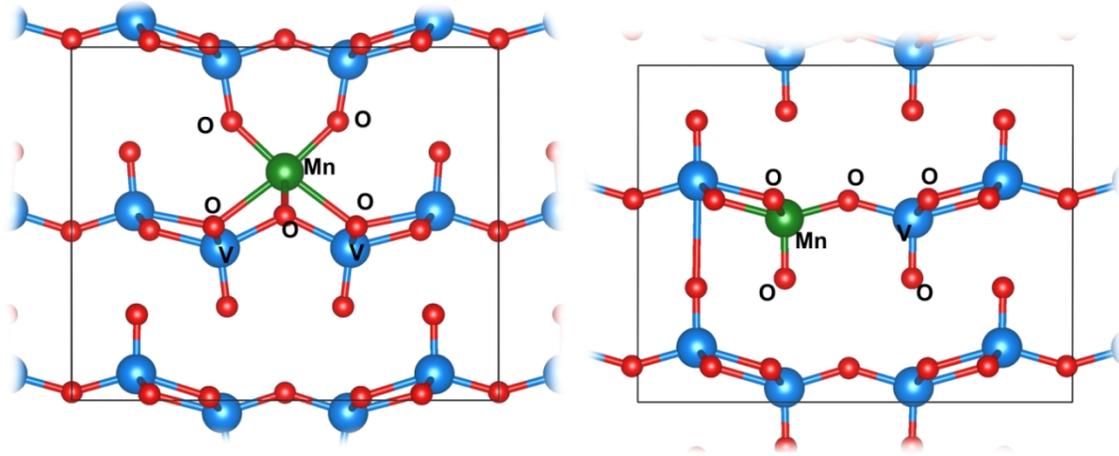

**Figure 5**. Unit cell of Mn doped $V_2O_5$. On the left the case of interstitial doping is shown while on the right substitutional doping is presented.

The volume changes and magnetization induced upon doping are presented in Table 2. As can be seen, interstitial doping induces an expansion of the unit cell, which is expected as the dopant atom is inserted between the $V_2O_5$ layers. The distance between the two layers accommodating the interstitial dopant increases as compared to the interlayer spacing without the dopant atom in the same system (Fig. 5). The effect of the interstitial impurity is seen for both layers either due to the insertion (pillaring) or due to the impact on the interactions between vanadyl oxygen atoms (O(1) atoms) and the V atoms of the subsequent layer (Fig. 4). In contrast, the unit cell volume of substitutionally doped $V_2O_5$ is only slightly affected, changing within ±1%. In general, the small volume changes obtained for doped $V_2O_5$ agree with the experimental findings where the orthorhombic structure of parental $V_2O_5$ is usually found to be preserved. For example, for Cu-doped $V_2O_5$ only the volume increase by 1.96% was reported but no change of the crystal structure.[14] It should be noted that in the case of doped $V_2O_5$ the addition of $U$ and D2 corrections has a similar effect as in the case of pristine $V_2O_5$. Namely, when the supercell volumes for the PBE and PBE+$U$ calculations are compared (not presented here), one can see that the latter are larger by roughly 5%. In contrast, when the D2 correction is added to PBE the supercell volume shrinks by approx. 10% for all the cases as a result of enhanced interlayer binding. The same also holds when the D2 correction is added to PBE+$U$.

**Table 2**. Change of the unit cell volume and total magnetization (*per* simulation cell) of doped $V_2O_5$ obtained by PBE+$U$+D2 calculations.

| Dopant | $\Delta V^*$ / % | | $M$ / $\mu_B$ | |
|---|---|---|---|---|
| | substitutional | interstitial | substitutional | interstitial |
| **Sc** | 0.81 | 5.25 | 2.00 | 1.00 |
| **Ti** | −0.75 | 3.74 | 1.00 | 2.00 |
| **Cr** | −0.25 | 3.52 | 1.00 | 6.00 |
| **Mn** | −0.60 | 3.48 | 2.00 | 5.00 |
| **Fe** | −0.31 | 3.91 | 3.00 | 4.00 |
| **Co** | −0.65 | 5.01 | 2.00 | 3.00 |
| **Ni** | 0.18 | 1.81 | 3.00 | 0.00 |
| **Cu** | 0.61 | 3.88 | 2.67 | 1.00 |
| **Zn** | 0.23 | 5.89 | 1.00 | 1.03 |

*evaluated as 100×($V_{\text{doped}}$ − $V_{\text{pristine}}$)/$V_{\text{pristine}}$; the volume of pristine 1×1×2 $V_2O_5$ supercell is 372.7 Å$^3$ using PBE+$U$+D2

In addition, an overall net magnetization arises upon the introduction of dopants into the $V_2O_5$ structure, and its value depends on the dopant type and the way the dopant is introduced into the lattice of $V_2O_5$ (substitutional/interstitial) (Table 2). In search for the origin of the magnetization we investigated the spin polarization densities (obtained as $\rho_{\text{spin up}} - \rho_{\text{spin down}}$) and found them to be located at the impurity atoms as well as on the surrounding V and O atoms. The spin polarization densities are presented in Fig. 6 for the case of Fe-doped $V_2O_5$. We note that the magnetization is found to be very sensitive to parameter $U$, and less sensitive to the D2 correction (as seen from the analogous PBE, PBE+$U$ and PBE+D2 calculations for all the systems). However, all these approaches agree with each other on the appearance of magnetization upon doping of $V_2O_5$ with 3$d$ elements.

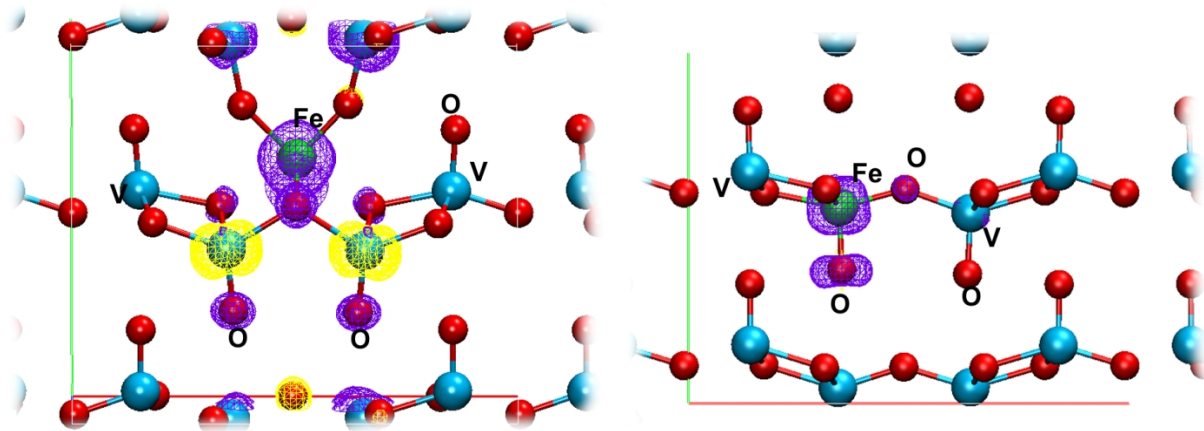

**Figure 6**. Distribution of spin polarization ($\rho_{\text{spin up}} - \rho_{\text{spin down}}$) for interstitially (left) and substitutionally (right) Fe-doped $V_2O_5$ (isovalues are ±0.0015 e Å$^{-3}$, positive isosurfaces are marked in purple color, while negative are yellow).

When it comes to the electrochemical applications of materials, conductivity plays a very important role as a hindered electron transport could limit the electrode performance. We investigated the electronic structures of doped $V_2O_5$ and found that, in general, the band gap becomes narrower due to the introduction of new dopant states (Fig. 7). This indicates that the conductivity of doped $V_2O_5$ should increase compared to parental $V_2O_5$ as it was indeed observed in some experimental reports.[16–21] In fact, by inspecting DOS curves (Fig. 7) one might see that practically a metallic behavior of otherwise insulating $V_2O_5$ is expected when doped with the elements between Mn and Ni. Therefore, it is clear that doping can be a powerful strategy for modifying the electrochemical and ion intercalation properties of $V_2O_5$. The conclusions derived from the analysis of DOS obtained using PBE+$U$+D2 hold also for the results obtained with PBE, PBE+$U$ and PBE+D2 calculations (not presented here for brevity).

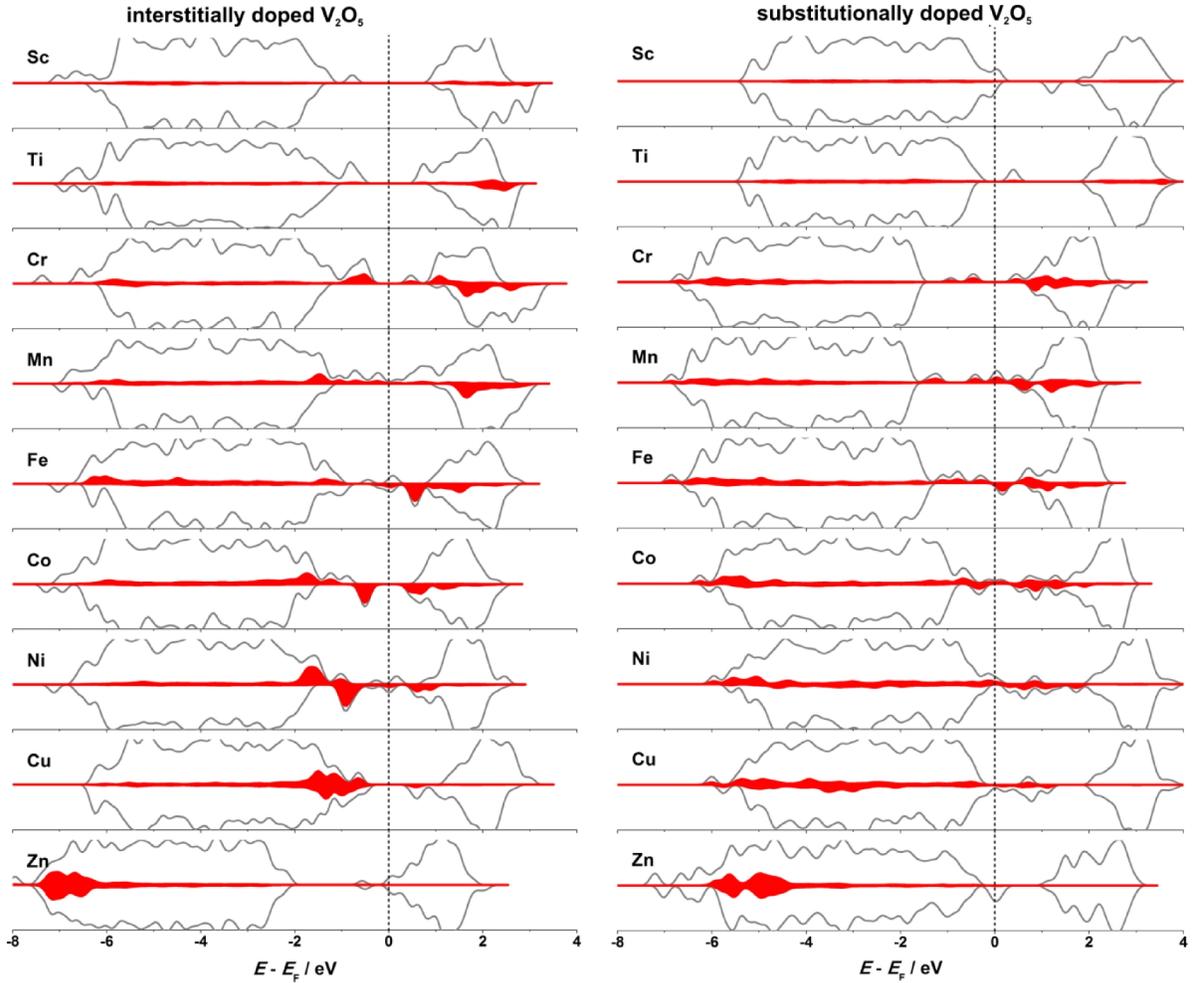

**Figure 7**. PBE+*U*+D2 calculated DOS of interstitially (left) and substitutionally (right) doped $V_2O_5$. Projected densities of *d*-states of dopant atoms (shaded) are also included.

In order to clearly identify the states that appear in the band gap of $V_2O_5$ upon doping, we analyzed the charge density distribution of these states (Fig. 8 for the cases of Mn-, Fe-, Co- and Ni-doped $V_2O_5$). We focused on the mentioned dopants causing nearly metallic behavior of doped $V_2O_5$. As expected from the DOS plots, the states located in the band gap of $V_2O_5$ are due to the dopant states and also the states of surrounding V and O atoms (Fig. 8).

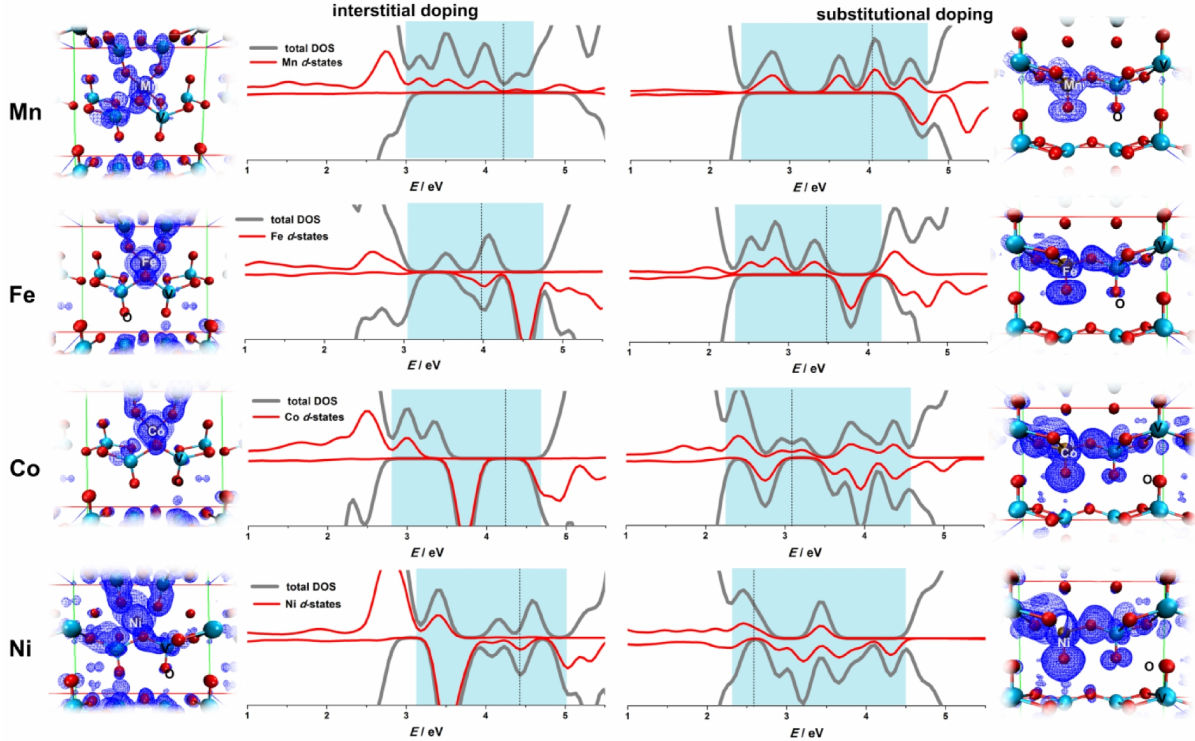

**Figure 8**. Projected density of states and the corresponding 3D charge density distribution maps of the states within the shaded energy window in the DOS plot. The results are presented for the case of $V_2O_5$ doping with Mn, Fe, Co and Ni (left side – interstitial doping, right side – substitutional doping, isosurfaces value is 0.002 e Å⁻³). Vertical dashed lines in the DOS plots indicate the Fermi levels.

These states are not highly localized and can indeed contribute to the conductivity of doped $V_2O_5$. It is difficult to derive some definite conclusions regarding the specific effects of interstitial *vs*. substitutional doping. For example, interstitially Co-doped $V_2O_5$ has a small band gap (~0.5 eV) while the substitutionally doped one shows a metallic behavior. At the same time Mn-, Fe- and Ni-doped $V_2O_5$ are metallic, irrespective of the type of doping. However, it is interesting to note that in the case of substitutionally doped $V_2O_5$ the states located in the band gap span within the layer containing the dopant, while the other layer is like in pristine $V_2O_5$, in terms of the electronic structure. On the other hand, in the case of interstitial doping the states are distributed between the layers, which host the impurity. This also indicates that the chemical interactions between the impurities and the V and O atoms of $V_2O_5$ lattice are more pronounced in the case of substitutional doping and this is also reflected in the overlap between the *d*-states of the impurity and the valence band of $V_2O_5$. For example, one can compare the cases of Co-, Ni- and Cu-doped $V_2O_5$ (Fig. 7) where it is clear that the *d*-states of the substitutional impurities span over the entire valence band of $V_2O_5$. In contrast, the *d*-states of interstitial impurities are predominantly located at the top of valence band. This also means that the chemical properties of interstitial and substitutional impurities, particularly in terms of the interactions with intercalated ions, will be rather different. Finally, a natural question, with no clear answer, remains – whether dopants will prefer substitutional of interstitial positions. Considering volume changes upon doping (Table 2) it seems that the elements from the first half of the 3$d$ row of Periodic Table are relatively easily incorporated into the $V_2O_5$ lattice. These elements can be found in many oxidation states (especially Cr, Mn) allowing them to replace $V^{5+}$ (this also applies to Co). In contrast, the elements with almost completed $d$-shell (Ni, Cu, Zn) have a limited number of oxidation states. Hence, we expect that, in reality, it would be difficult for these atoms to adapt to the coordination of $V^{5+}$ in $V_2O_5$ structure. Naturally, we expect that the preference between substitutional and interstitial doping will also depend on the concentration of dopants. For example, Guan $et$ $al.$[22] described preparation and electrochemical properties of Mn-doped $V_2O_5$ in a very wide range of concentrations, from $Mn_{0.27}V_2O_5$ (which is very close to our model) to $Mn_{0.94}V_2O_5$. For low Mn concentrations the XRD patterns of Mn-doped $V_2O_5$ corresponded to that of pure $V_2O_5$ while for high concentrations of Mn the $V_2O_5$ interlayer spacing was found to significantly increase. When this result was combined with different types of V=O bonds evidenced in Raman spectra, it was concluded that Mn was inserted between $V_2O_5$ layers and distorted the $V_2O_5$ structure. Such different V=O bonds are clearly seen in Fig. 5 for the case of interstitially Mn-doped $V_2O_5$ (for the analysis of the effect of addition of $U$ on Mn 3$d$ states on the structure of Mn-doped $V_2O_5$ reader is referred to Supplementary Information). The authors concluded that the structural changes of $V_2O_5$, being the consequence of doping, provided more free space for the $Li^+$ intercalation/de-intercalation. Unfortunately, the same report contains no information about the conductivity of doped $V_2O_5$.

## 4. Conclusions

By employing the combination of the DFT+$U$ approach with the semi-empirical DFT+D2 correction for the long-range dispersion interactions, both the crystal and the electronic structure of $V_2O_5$ were adequately described using periodic plane wave DFT calculations. The inclusion of the D2 correction is of primary importance to address the interlayer spacing in bulk $V_2O_5$ while the intralayer chemical bonding is not significantly affected by the addition of the $U$ term and D2 correction. Within the PBE+$U$+D2 scheme computational costs are increased by roughly 30-40%, which is solely ascribed to the calculation of the $U$ correction. Using the same approach, we investigated the effects of doping of $V_2O_5$ by 3$d$ elements, both interstitially and substitutionally. Interstitial doping was found to result in an expansion of the $V_2O_5$ lattice, while substitutional doping has a much less pronounced effect on the structure of parental $V_2O_5$. However, doping induces significant changes of the electronic structure and leads to a narrowing of the band gap of $V_2O_5$. This is expected to result in somewhat higher conductivity of doped $V_2O_5$, which was indeed observed in some of the previously published experimental studies. The obtained results suggest that doping can be an elegant strategy for modifying structural and electronic properties of $V_2O_5$. This is of crucial importance for the application of this material in the field of metal-ion batteries.

## Conflicts of interest

There are no conflicts to declare.

## Acknowledgement

This work was supported by the Serbian Ministry of Education, Science, and Technological Development (III45014). S.V.M is indebted to Serbian Academy of Sciences and Arts for funding this study through the project "Electrocatalysis in the contemporary processes of energy conversion". N.V.S. acknowledges the support provided by Swedish Research Council through the project No. 2014-5993. This work was additionally supported by the COMET program by the Austrian Research Promotion Agency (FFG) and the governments of Lower and Upper Austria. We also acknowledge the support from Carl Tryggers Foundation for Scientific Research. The computations were performed on resources provided by the Swedish National Infrastructure for Computing (SNIC) at the High Performance Computing Center North (HPC2N) at Umeå University.

# SUPPLEMENTARY INFORMATION

**The effects of inclusion of +$U$ correction on dopant atoms – Cases of Mn- and Co-doped V$_2$O$_5$**

We present here the overview of the results of the calculations where +$U$ correction was applied also onto the dopant 3$d$ states. In order to demonstrate the effects we chose the cases of Mn and Co. Following the overview of Capdevila-Cortada *et al*.[S1] a wide range of values of $U$ were applied so far for these two elements. For CoO$_x$ compounds $U$ was found in the range 3.3 to 6.7 eV, and most frequently was chosen by fitting experimental properties. For MnO$_x$ compounds $U$ was found in the range 1 to 6.63 eV. Here we see that the addition of $U$ on dopants does not affect overall conclusions regarding the expansion of lattice (Table S1). The changes are much smaller in the case of substitutional doping. Also, an increase of the value of $U$ applied onto dopant 3$d$ states in general leads to the expansion of the lattice (Table S1).

**Table S1.** Change of the unit cell volume ($\Delta V$ / %)* of Mn- and Co-doped V$_2$O$_5$ obtained by PBE+$U$+D2 calculations. The values of $U$ applied on dopant 3$d$ states was varied while the value of U applied to V 3$d$ states was kept to 6 eV.

| $U$@M / eV | Mn-doped V$_2$O$_5$ | | Co-doped V$_2$O$_5$ | |
|---|---|---|---|---|
| | **substitutional** | **interstitial** | **substitutional** | **interstitial** |
| **0** | -0.60 | 3.48 | -0.64 | 5.01 |
| **2** | -0.48 | 4.14 | -0.55 | 5.43 |
| **4** | -0.11 | 8.30 | -0.52 | 2.25 |
| **6** | 0.57 | 8.98 | 0.67 | 6.37 |

*evaluated as 100×($V_{doped} - V_{pristine}$)/$V_{pristine}$; the volume of pristine 1×1×2 V$_2$O$_5$ supercell is 372.7 Å$^3$ using PBE+$U$+D2

Here we also show the electronic structure of Co-doped V$_2$O$_5$ (Fig. S1). It can be seen that the effects of addition of $U$ on Co 3$d$ states (in addition to the V 3$d$ states) affects the electronic structure, as expected. The case of interstitial doping is more sensitive to the addition of $U$ on dopant states. In the case of substitutional doping we see that the band gap of parental V$_2$O$_5$ is completely lost, irrespectively on the value of $U$ applied to the Co 3$d$ states. Nevertheless, without proper experimental reference it is difficult to derive a definite conclusion regarding the addition of $U$ to dopant states.

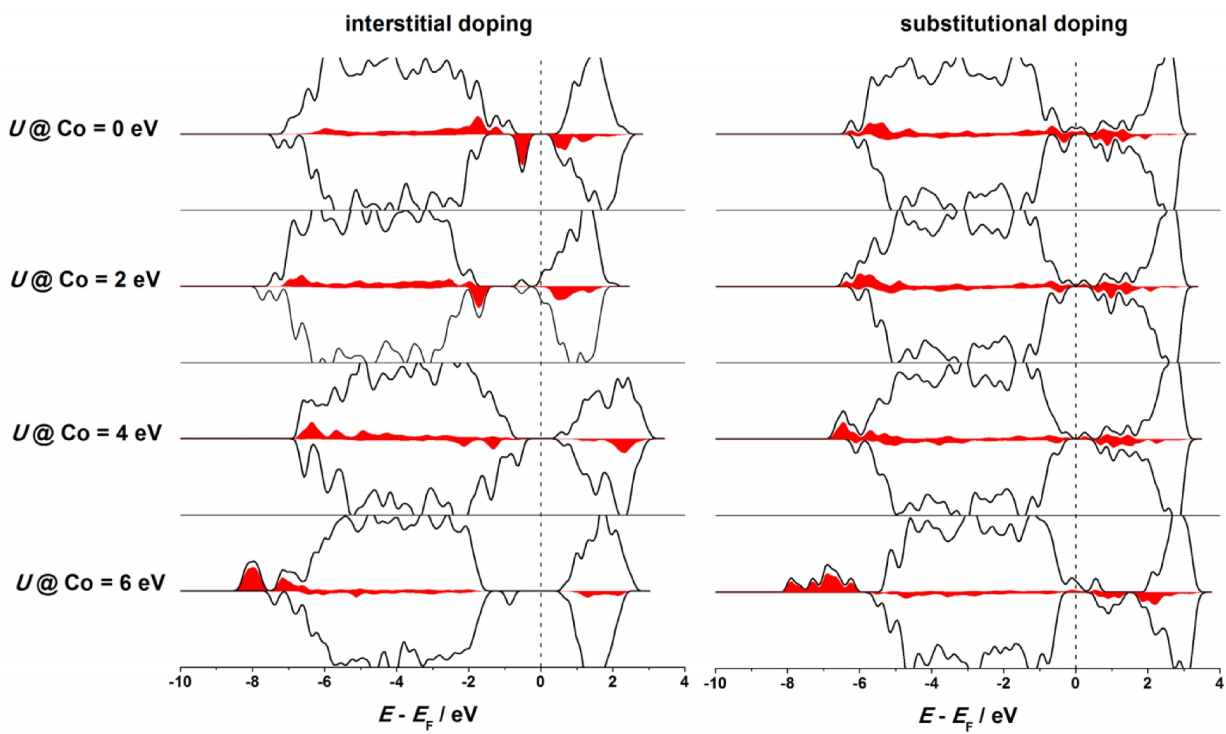

**Figure S1**. DOS plots for Co-doped $V_2O_5$ (projected densities of dopant states are shaded) obtained by PBE+$U$+D2 calculations. The values of $U$ applied on dopant $3d$ states were varied while the value of U applied to V $3d$ states was kept to 6 eV. Vertical dashed lines denote Fermi levels.

## Supplementary references